# Deep Electron Cloud-activity and Field-activity Relationships

Lu Xu [a,d]* and Qin Yang [b,c]*

[a] *School of Sports and Health Science, Tongren University, Tongren 554300, PR China*

[b] *School of Physics and Optoelectronic Engineering, Yangtze University, Jingzhou 434023, PR China*

[c] *Neonatal Screening Center, Shanghai Children's Hospital, Shanghai Jiao Tong University, Shanghai, 200040, PR China*

[d] *College of Material and Chemical Engineering, Tongren University, Tongren 554300, PR China*

*Corresponding authors. E-mail addresses: yqbioinformatics@126.com (Qin Yang); lxchemo@163.com (Lu Xu).

**Abstract**

Chemists have been pursuing the general mathematical laws to explain and predict molecular properties for a long time. However, most of the traditional quantitative structure-activity relationship (QSAR) models have limited application domains, e.g., they tend to have poor generalization performance when applied to molecules with parent structures different from those of the trained molecules. This paper attempts to develop a new QSAR method that is theoretically possible to predict various properties of molecules with diverse structures. The proposed deep electron cloud-activity relationships (DECAR) and deep field-activity relationships (DFAR) methods consist of three essentials: (1) A large number of molecule entities with activity data as training objects and responses; (2) three-dimensional electron cloud density (ECD) or related field data by the accurate density functional theory methods as input descriptors; (3) a deep learning model that is sufficiently flexible and powerful to learn the large data described above. DECAR and DFAR are used to distinguish 977 sweet and 1965 non-sweet molecules (with 6-fold data augmentation) and the classification performance is demonstrated to be significantly better than the traditional least squares support vector machine (LS-SVM) models using traditional descriptors. DECAR and DFAR would provide a possible way to establish a widely applicable, cumulative, and shareable artificial intelligence-driven QSAR system. They are likely to promote the development of an interactive platform to collect and share the accurate ECD and field data of millions of molecules with annotated activities. With enough input data, we envision the appearance of several deep networks trained for various molecular activities. Finally, we could anticipate a single DECAR or DFAR network to learn and infer various properties of interest for chemical molecules, which will become an open and shared learning and inference tool for chemists.

# 1. INTRODUCTION

Chemists have been pursuing the general mathematical laws or rules to explain and predict molecular properties for a long time[1,2]. Unfortunately, at least until today, it seems to be still a distant dream to predict the various properties of interest for millions of chemical molecules with simple and widely applicable principles[3]. Although quantum mechanics or quantum chemistry provides a fundamental and powerful tool to reveal and explain molecular properties, there are still some difficulties in how to make good use of quantum chemistry and explain its results[4]. For instance, the complexity of the whole action system (including a drug molecule, media, targets and pathways) is usually large and computationally too heavy for today's strict quantum chemical methods and computational resources[5,6]. Moreover, quantum mechanics generally calculates those properties derived from or defined on clear physical actions or relationships. However, many molecular properties that are useful and of big interest to humans, such as the activity of a chemical molecule against cancer cells, are obviously caused by multiple and very complex physical or/and chemical actions and defined based on human experience or/and experiments[7]. Prediction of the molecular properties defined by human experience is precisely the area where machine learning and inference are applicable and have achieved great success[8,9].

Quantitative structure-activity relationship (QSAR) methods using machine learning has been applied in various fields, such as chemistry, pharmacology, biology, materials, and environmental sciences[10-14]. However, many traditional QSAR models are based on small data sets including hundreds or even tens of molecules. Although they can sometimes obtain good predictions and interpretability, their application range is limited and they usually perform poorly in predicting the properties of molecules with diverse parent structures[15]. Moreover, the screening of large data sets is mostly based on some empirical rules or inaccurate descriptors with approximate information on molecular structure[16]. To date, thousands of traditional molecular descriptors have been designed to allow computers to process chemical molecules[17]. However, more or less, most of them are mixed with experience and approximation, which are insufficient for accurately reflecting the information carried by a real molecular structure[18]. Therefore, the traditional descriptors and QSAR methods cannot guarantee the reliability and generalization performance, nor can they obtain accurate and widely applicable mathematical laws or machine learning models.

To develop a QSAR model that is possible to predict the properties for diverse molecular structures, obviously, the following requirements need to be met:

(1) A large number of molecule entities (objects with responses or labels) should be used as input. One of chemists' core tasks is to test and discover useful molecular properties in different fields. Chemists have synthesized and discovered millions of molecules, and our knowledge of molecular activities is growing rapidly every day. Therefore, we will have more and more molecules with annotated activity data for learning.

(2) Strict and universal molecular descriptors (features) should be generated to represent the full and accurate molecular structure information. According to the Hohenberg-Kohn theorem[19], the three-dimensional (3D) electron cloud density (ECD) of a molecule determines all the electronic properties of its ground state. Therefore, this paper will study the feasibility of using the 3D ECD data or the related field data (such as electrostatic potential field (ESPF) data) derived from ECD as the molecular descriptors.

(3) A sufficiently powerful and flexible model to learn from the massive molecular information. Fortunately, deep learning (with multi-hidden-layer neural networks) has provided us with a powerful tool for learning from large data sets[20, 21]. Deep learning has achieved unprecedented success in many fields of artificial intelligence (AI), especially image analysis and classification of 3D point clouds[22, 23], where the learning ability of big data and its advantages in capturing local details using convolutional techniques have been brilliant.

The aim of this paper is to design a reliable and widely applicable QSAR method. Based on the above considerations, we propose two new methods, study their feasibility and capability with real data, and compare their performance with some traditional QSAR methods.

## 2. METHODS

### 2.1 Definition of deep electron cloud-activity and field-activity relationships

To obtain QSAR models that are possibly applicable to diverse molecular structures, we propose using multilayer 3D convolutional neural networks (3D-CNNs) to perform deep learning of molecular physicochemical properties based on the 3D ECD data (or the spatial field data derived from ECD) of a large number of chemical molecules. We refer to the proposed methods as deep electronic cloud-activity relationships (DECAR) and deep field-activity relationships (DFAR). The three essential requirements of DECAR and DFAR include the following:

(1) Only by learning from a large number of molecules can a QSAR model theoretically predict the properties of diverse molecular structures. Therefore, the first essential requirement is a large number of chemical molecules (with class labels or annotated activity values) as input objects, from thousands to tens of thousands or even hundreds of thousands or millions, which may approach the limit of our increasing computing resources and capacity.

(2) Full and accurate molecular structural information should be extracted as input features. According to the Hohenberg-Kohn theorem, the 3D ECD of a molecule determines all the electronic properties of its ground state. Therefore, we propose to use 3D ECD data or related field data derived from calculations by the accurate density functional theory (DFT) methods as molecular structure descriptors. That is, the second essential requirement is the high-quality 3D ECD or field data of chemical molecules as input features.

(3) The third essential requirement of DECAR or DFAR is a flexible and powerful machine learning method that can deal with big data. In this paper, deep learning with 3D-CNNs is used because it can learn a large amount of molecular spatial structure information and the model can be constantly improved with the increase of effective training data. CNNs can capture the global and local details of 3D ECD or field data, which determines the various properties of a molecule. Moreover, deep learning of big augmented data can overcome the problem of feature alignment caused by the translation and rotation (or orientation) of molecules in a 3D space[24,25].

## 2.2 The network structures

A multilayer 3D-CNN is proposed to learn molecular spatial structures represented by 3D ECD or ESPF data. The network mainly consists of 3 convolutional layers, 3 max-pooling layers, and 2 fully connected layers (Figure 1). Three-dimensional ECD or ESPF data (in arrays of 200×200×200 points) are fed into the first layer as the input. Each 3D convolutional layer is followed by batch normalization, leaky rectified linear unit (ReLU) and a 3D max-pooling operation. Convolutional layers convolved the learned filters with the input data and produced feature maps for each filter, extracting global-local features of the molecular spatial structure. The kernel sizes of all convolutional layers are 8×8×8 with a 1×1×1 stride, and their filter numbers are set to 16, 32, and 64. The following pooling layers reduce the size of the feature map by replacing each cube with their maximum value and keep the most influential features for discriminating molecular structures. Max pooling is applied to each 3×3×3 cube with a 1×1×1 stride. A ReLU is employed as the activation function in all convolutional and pooling layers. The last layers are two fully connected layers to perform the final classification and obtain the output. They consist of a number of input and output neurons that generate a learned linear combination of all neurons from the previous layer and pass it through a nonlinear function. The numbers of nodes in the fully connected layers are 64 and 2. Softmax is used as the activation function connecting the fully connected layer and the output layer to obtain probability values for classification.

[Please insert Figure 1 here]

The 3D-CNN model is trained using the adaptive moment estimation (Adam) optimizer with the recommended parameters $\beta_1 = 0.9$, $\beta_2 = 0.999$, $\epsilon = 10^{-8}$ and a weight decay rate of 0.01. One cycle learning rate scheduling is used as well, in which the learning rate is initially set to a slightly lower value and periodically increased during the first 30% of batches. After reaching the maximum, the learning rate is slowly annealed until the end. In the training process, the cross-entropy function is used as the loss function of the network. Batch normalization and L2 regularization (with a regularization factor of 0.0005) are applied to improve the generalization performance and avoid problems such as vanishing gradients and overfitting. The maximum number of epochs is set to 100, and the experimental batch size is set to 20 in each iteration. Additionally, the order of the training data is shuffled in every epoch.

**2.3 Data of sweet and non-sweet molecule entities**

A total of 2942 molecular tastants are collected from ChemTastesDB[26]. Each molecule is classified into one of the five basic tastes (sweet, bitter, umami, sour and salty), as well as into four non-basic classes, such as tasteless, non-sweet, multitaste and miscellaneous. ChemTastesDB provides the following information for each molecule: name, PubChem CID, CAS registry number, canonical SMILES string, class taste and references to the scientific sources from which the data were retrieved. Moreover, the molecular structure of each chemical is provided. In this study, the aim is to discriminate the 977 sweet and 1965 non-sweet molecules according to their molecular structures. Each molecule is labeled sweet or non-sweet.

**2.4 Computation of molecular structural features**

The geometric structure of each molecule is optimized by Gaussian 16 (Gaussian, Inc., Wallingford, USA) using the B3LYP method with 6-31G (d, p) basis sets. ECD and ESPF data are calculated with the Multiwfn software[27]. When sampling the ECD and ESPF data, the molecule cannot be zoomed in and out like an image because the properties and functions of a molecule depend on both the spatial distributions of its ECD (or ESPF) and its size. In other words, if the structures of two ECDs are geometrically identical or similar but of different sizes, they would have different molecular properties. Therefore, it is necessary to use grids of uniform size to generate the ECD or ESPF data for all the molecules. Given a molecule, if there are too many grids, the size of the data array will be very large; if the number of mesh points is too small, the sampling accuracy will be degraded. In this paper, considering the size range of all the 2942 molecules, a 200×200×200 cubic array is generated for each molecule, and the real interval between grid points is 0.5 Bohr, which balances the final data size and the sampling accuracy. The molecular structural formula and 3D ECD array of a typical sweet molecule, D-glucose, is shown in Figure 2.

[Please insert Figure 2 here]

## 2.5 Data augmentation and division

Since a molecular structure generally does not have symmetry in the 3D space, actually there are countless orientations for a molecule, so the selection of representative orientations of the data is very important. In this paper, the ECD and ESPF data are generated as a cube. If each of the six faces of the cube faces upward once, each molecule can generate six representative data arrays. In this way, for the same molecule, any two augmented data arrays must be rotated by at least 90 degrees to coincide with each other.

The above data augmentation will make the final prediction step to be slightly different. Each test object has six rotated data arrays, so each object is predicted six times. For each prediction result, we use softmax to calculate the probabilities that it belongs to the two classes as shown in Eq. (1). Then, the six probabilities were summed up to obtain the total probabilities that a molecule belongs to the two categories as shown in Eq. (2). In this way, the final classification of the object is achieved.

$$p_{ij} = \text{softmax}(x)_{ij} = \frac{e^{x_{ij}}}{\sum_{i=1}^{2} e^{x_{ij}}} \quad (i=1\sim2, j=1\sim6) \quad (1)$$

$$p_i = \sum_{j=1}^{6} p_{ij} \quad (i=1\sim2, j=1\sim6) \quad (2)$$

where $p_{ij}$ is the probability that a molecule belongs to class $i$ predicted according to the $j$th data array of that molecule and $p_i$ is the probability that a molecule belongs to class $i$ predicted according to all the six data arrays of that molecule.

For the 2942 molecular entities, the original data are randomly split into three subsets: 70% (2060 molecules, 684 sweet and 1376 non-sweet) for training, 15% for validation (441 molecules, 147 sweet and 294 non-sweet) and 15% for testing (441 molecules, 146 sweet and 295 non-sweet). Since each molecule has 6-fold augmented data arrays, the final training set contains 12360 data, and the validation and test sets each contain 2646 data.

## 2.6 Traditional machine learning methods

Meanwhile, traditional QSAR models based on least squares support vector machine (LS-SVM) [28,29] were also developed for comparison using the same molecule partition (2060 molecules for training, 441 for validation and 441 for inference). Based on the molecular geometry optimized by Gaussian 16, Dragon 7 software (Kode SRL, Milano, Italy) was used to calculate the values of 5270 traditional molecular descriptors. Particle swarm optimization (PSO) algorithm[30] was used to perform feature selection and parameter optimization of LS-SVM. Different kernel functions, including the linear kernel and the nonlinear radial basis function (RBF) kernel were used by LS-SVM.

**2.7 Softwares and computing resources**

Classification models are developed on MATLAB R2022a (MathWorks, Sherborn, MA, USA). The 3D-CNN is trained using the Deep Learning toolbox of MATLAB. The PSO-LS-SVM algorithm is developed using the LS-SVMlab v1.8 toolbox[31] with self-compiled MATLAB codes for PSO. Heatmap is plotted using the package pheatmap (v1.0.12) in R (v4.1.0). Computing resources include 6 servers, each with two Intel Xeon E5-2680 V4 CPUs and 128 GB RAM (for running Gaussian 16), as well as another server with a series of 8 NVIDIA Tesla P40 24G GPUs and 128 GB RAM (for deep learning).

**3. RESULTS AND DISCUSSION**

For PSO-LS-SVM with traditional chemical descriptors, PSO is performed to select a subset of the 5270 features and optimize the model parameters of LS-SVM. Considering the sizes of the training and validation sets, the following objective function of weighted errors (WE) is minimized to balance the quality of training and validation:

$$WE = \sqrt{\frac{SST}{2060} + \frac{SSV}{441}} \quad (3)$$

where $SST$ is the sum of the squared residuals of the training set, $SSV$ is the sum of the squared residuals of the validation set, and 2060 and 441 are the sizes of the training set and validation set, respectively.

[Please insert Figure 3 here]

Learning curves of DECAR (Figure 3) is plotted to demonstrate the training process of neural networks. By examining the curves of accuracy and cross entropy loss, the improvement of the network is relatively fast in the first few cycles, then gradually slows down, and finally converges to the best result. Table 1 summarizes the classification results of PSO-LS-SVM, DECAR and DFAR. In general, the 3D-CNN models outperform the LS-SVM methods with traditional chemical descriptors. PSO-LSSVM with the RBF kernel achieves a prediction accuracy of 0.8277, a sensitivity of 0.8014, and a specificity of 0.8407. Based on ECD and ESPF rather than traditional molecular descriptors, both DECAR and DFAR show great superiority in classification performance. On the test set, the accuracies are 0.9388 and 0.9433, the sensitivities are 0.8904 and 0.9178, and the specificities are 0.9627 and 0.9559 for DECAR and DFAR, respectively. In addition, as 3D-CNN has much more flexibility than LS-SVM and considering their performances in image classification, DECAR and DFAR are predicted to obtain further improved classification results with more molecules used for training. In contrast, traditional methods such as LS-SVM do not show obvious advantages on large datasets.

[Please insert Table 1 here]

Why do DECAR and DFAR achieve significantly better classification accuracy than the traditional PSO-LS-SVM method? Obviously, the results should be mainly attributed to the powerful and flexible modeling ability of deep learning, as well as the sufficiency of ECD or ESPF data as molecular features. According to the Hohenberg-Kohn theorem, the 3D ECD of a molecule determines all the electronic properties of the molecule in its ground state. Therefore, 3D ECD or related field data can provide more accurate and reliable information for modeling and inferring molecular properties. Then, how does the 3D-CNN capture the structural information that determines whether a molecule has sweetness? We try to explain it based on the maximum correlation between convolution kernels and a molecule. For all convolution kernels of each convolution layer, we calculate the maximum normalized convolution (MNC) with all molecules:

$$\text{MNC}_{ijk} = \max\left(r(\mathbf{CK}_{ij}, \mathbf{X}_{klm})\right) \qquad (4)$$

Where $\mathbf{CK}_{ij}$ is the *i*th convolution kernel in the *j*th convolution layer and $\mathbf{X}_{klm}$ is the *l*th sub-block of the *m*th rotation of the *k*th object (molecule), which has the same size with $\mathbf{CK}_{ij}$. Person's correlation coefficient (*r*) is computed between vectorized $\mathbf{CK}_{ij}$ and $\mathbf{X}_{klm}$ over a whole molecule with a stride of 1×1×1. $MNC_{ijk}$ can reflect the maximum local similarity between a convolution kernel and a molecular data array.

[Please insert Figure 4 here]

The heatmap of MNC (Figure 4) shows the evolution of the maximum local similarity between convolution kernels and molecules. For all the three convolution layers, the MNC values of the 977 sweet molecules are generally higher than those of the 1965 non-sweet molecules. The clustering results also indicate the ability of the convolution kernel to distinguish sweet and non-sweet molecules. Interestingly, both the difference in MNC visible to the naked eye and the ability of MNC to distinguish the two classes of molecules (by clustering) become more prominent with the deepening of the convolution. This shows that the evolution of multilayer convolution kernels is very helpful for learning the local characteristics of sweet and non-sweet molecules.

4. **Conclusion**

To develop a QSAR method that is possibly applicable to diverse molecular structures and activities, this paper put forward DECAR and DFAR and studied their feasibility and capacity with actual data. In discriminating the sweet and non-sweet molecules, DECAR and DFAR both showed better learning and inference abilities than traditional LS-SVM models. Based on the results, we would draw the conclusions of this paper.

(1) By combining deep learning, 3D ECD data (or related field data) as input descriptors, and a large number of chemical molecules, DECAR and DFAR would provide a possible QSAR approach and the application domain could theoretically cover diverse chemical structures and various molecular properties. According to the Hohenberg-Kohn theorem, ECD and related field data by DFT indeed provide the most accurate and abundant molecular structure information to date, so they have the potential to serve as strict and universal descriptors for chemical molecules. Based on the great success of deep learning in classification of images and point clouds, as well as the results obtained by this paper, we are optimistic that DECAR and DFAR may provide a possible method to achieve widely applicable QSAR, which is worthy of in-depth and extensive research. Especially, further research should be focused on larger number of molecular entities and many other molecular properties or activities besides sweetness.

(2) The huge demand of DECAR and DFAR for data accumulation and computing would require the development of a cumulative and shareable AI-driven QSAR system. To date, millions of chemical molecules have been discovered and designed and many of their properties or activities are unknown and need to be inferred and tested. One of our future work will involve building an international platform for collecting and sharing the ECD and field data for millions of chemical molecules. That might be a molecule "ImageBank" of chemists and could be developed in an interactive way like Wikipedia, where individuals can prepare, upload, share, and revise their own molecular ECD data (or field data). The ECD data could be annotated with various known activity data in a similar way. Once a large amount of molecular data and information has been accumulated, more networks of various molecular activities could be trained and shared, which would be useful for inference and understanding of molecular properties.

(3) Finally, we will conclude this article with a further outlook of DECAR and DFAR. Recognizing the fact that the AlexNet has learned and classified 1000 classes of images[32], with enough inputs of molecule entities and activity data, could we anticipate a single network that can learn and infer various (hundreds or thousands of) properties of interest for diverse molecular structures in the future? We think that this might be possible.

**ACKNOWLEDGEMENTS**

This work was supported by funding from National Natural Science Foundation of China (Nos. 82260896 and 21803009), the Science Foundation of Educational Commission of Hubei Province of China (No. T2020008), Guizhou Provincial Science and Technology Department (No. QKHPTRC[2020]5009), and Tongren Science and Technology Bureau (No. TSKY2019-3).

**CONFLICT OF INTEREST**

**ETHICS STATEMENT**

Not applicable.

**DATA AVAILABILITY STATEMENT**

Data available on request from the authors.

**ORCID**

Lu Xu: https://orcid.org/0000-0003-4742-5623

Qin Yang: https://orcid.org/0000-0002-3773-4169

**Figure and table captions**

Table 1. Classification results of PSO-LSSVM, DECAR and DFAR

Figure 1. Network structure of DECAR and DFAR

Figure 2. The molecular structural formula (a) and 3D ECD (b) of D-Glucose

Figure 3. Learning curves of DECAR for accuracy (a) and cross entropy loss (b)

Figure 4. Heatmap of MNC between convolution kernels and sweet/non-sweet molecules (a~c correspond to convolution layers 1~3)

**Table 1.** Classification results of PSO-LSSVM, DECAR and DFAR

| Model | Parameters [a] | Training set (684+1376) | Validation set (147+294) | Test set (146+295) | | |
|---|---|---|---|---|---|---|
| | | Accuracy /Loss [b] | Accuracy /Loss | Accuracy /Loss | Sensitivity | Specificity |
| PSO-LSSVM (linear kernel) | 89 descriptors, $\gamma$ =6.97 | 0.8034 | 0.7937 | 0.7755 | 0.7466 | 0.7898 |
| PSO-LSSVM (RBF kernel) | 136 descriptors, $\sigma^2$=630, $\gamma$ =15.33 | 0.8660 | 0.8735 | 0.8277 | 0.8014 | 0.8407 |
| DECAR | - | 0.9791/ 0.0086 | 0.9546/ 0.0235 | 0.9388 /0.0342 | 0.8904 | 0.9627 |
| DFAR | - | 0.9854/ 0.0046 | 0.9615/ 0.0248 | 0.9433/ 0.0307 | 0.9178 | 0.9559 |

[a] $\sigma^2$ is the kernel width of RBF and $\gamma$ is the regularization parameter.

[b] Cross entropy loss for deep learning.

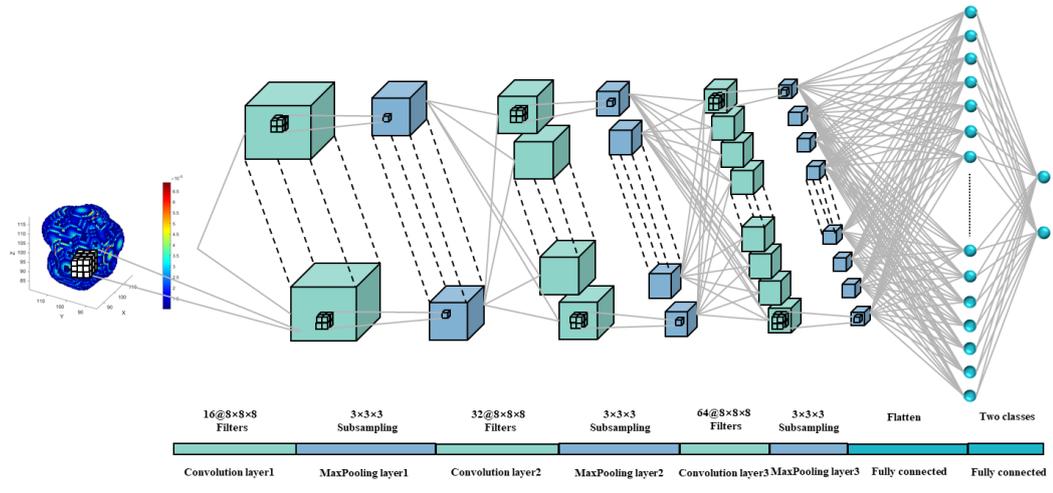

Figure 1. Network structure of DECAR and DFAR

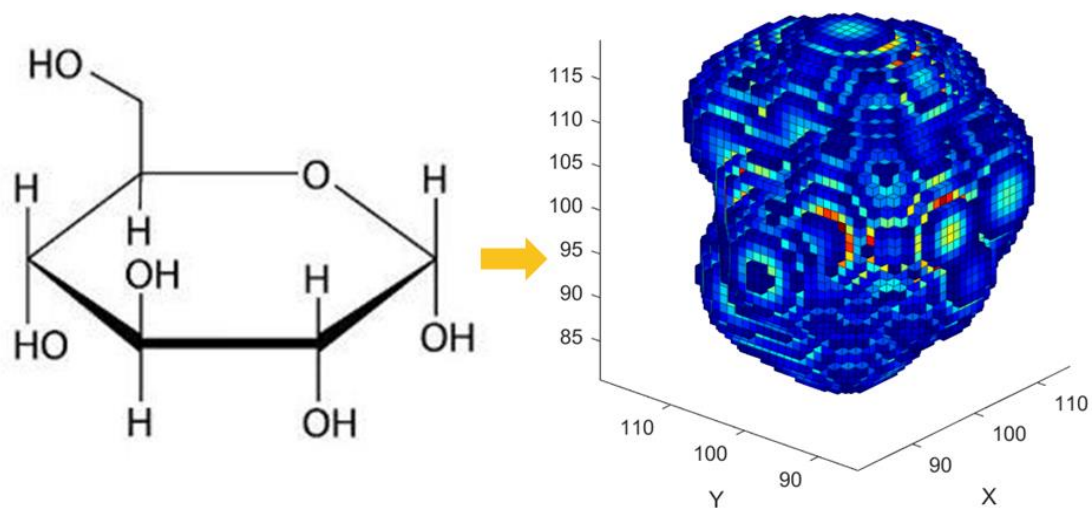

Figure 2. The molecular structural formula (a) and 3D ECD (b) of D-Glucose

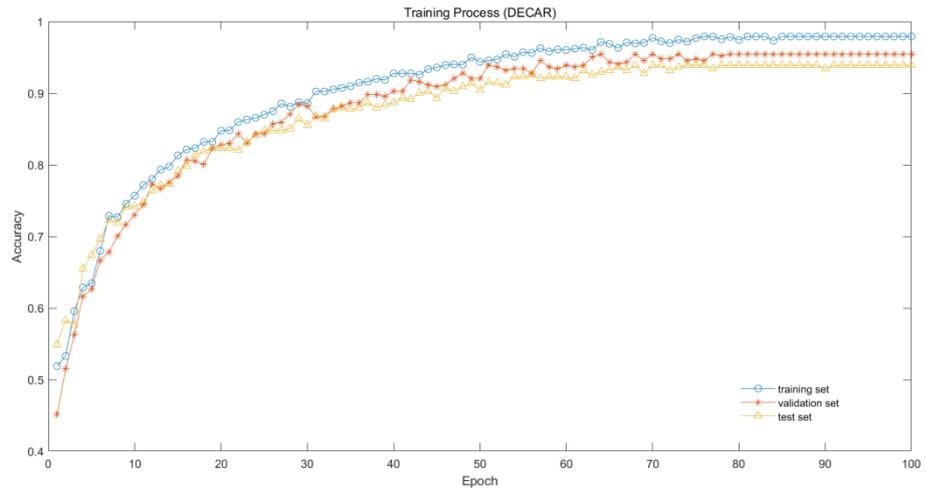

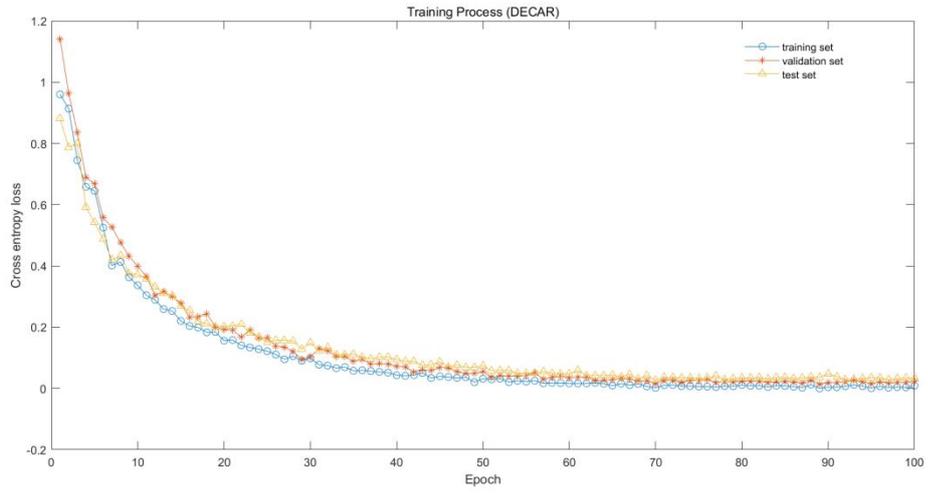

Figure 3. Learning curves of DECAR for accuracy (a) and cross entropy loss (b)

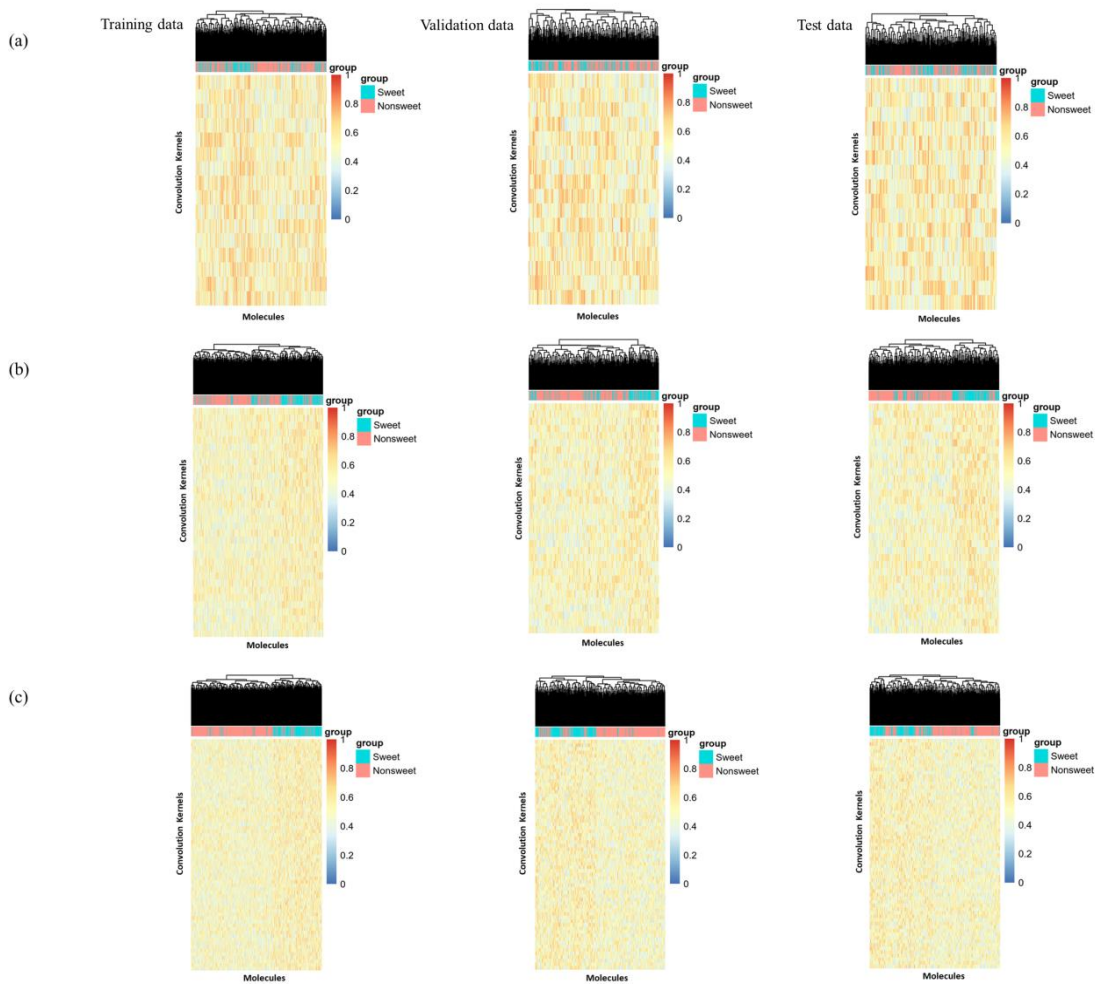

Figure 4. Heatmap of MNC between convolution kernels and sweet/non-sweet molecules (a~c correspond to convolution layers 1~3)